\documentclass[parskip=half, bibtotoc]{scrartcl}

\usepackage[utf8]{inputenc}
\usepackage[T1]{fontenc}
\usepackage[ngerman,english]{babel}

\usepackage{authblk} 
    
\usepackage{csquotes}
\usepackage[breaklinks=true]{hyperref}
\usepackage[most]{tcolorbox}
\usepackage{wrapfig}
\usepackage{xurl}
\usepackage{fontawesome5}

\usepackage[style=numeric, backend=biber, sorting=none]{biblatex}

\usepackage{amsmath}
\usepackage{amssymb}
\usepackage{amsfonts}
\usepackage{mathtools}
    \mathtoolsset{showonlyrefs=true}
    
\usepackage{xspace}
\usepackage{graphicx}
\usepackage{xcolor}
\usepackage{orcidlink}

\usepackage[lang-warn=false]{datatool-base}  
\usepackage[acronym]{glossaries}
\usepackage{xstring}

\usepackage{enumitem}

\providecommand*{\eg}{e.g.\@\xspace}

\providecommand*{\cf}{cf.\@\xspace}

\newcommand{\package}[1]{\texttt{#1}\@\xspace}
\newcommand{\tool}[1]{\textit{#1}\@\xspace}
\newcommand{\code}[1]{\texttt{#1}}
\DeclareEmphSequence{\bfseries}

\renewcommand{\abstract}[1]{\hfill\begin{minipage}{0.9\linewidth}%
                                \phantomsection\pdfbookmark[2]{Abstract}{abstract}%
                                \textbf{Abstract: }#1
                            \end{minipage}\hfill}
\providecommand{\keywords}[1]{\phantomsection\pdfbookmark[2]{Keywords}{keywords}\textbf{Keywords: }#1}
\providecommand{\acknowledgement}[1]{\phantomsection\pdfbookmark[2]{Acknowledgement}{acknowledgement}\textbf{Acknowledgement: }#1}

\AtBeginDocument{
    \phantomsection\pdfbookmark[1]{Title}{title}
}

\makeatletter 
    \newcommand{\myhypertarget}[1]{\Hy@raisedlink{\hypertarget{#1}{}}}
\makeatother
\newcommand\createacronym[3]{%
    \newacronym{#1}{#2}{#3}%
    \StrDel{#2}{-}[\myString]%
    \StrDel{#2}{/}[\myString]%
    \expandafter\newcommand\csname \myString\endcsname{%
        \hyperlink{#1}{\gls*{#1}}%
        \myhypertarget{#1}{}\xspace%
    }%
    \expandafter\newcommand\csname \myString s\endcsname{%
        \hyperlink{#1}{\glspl*{#1}}%
        \myhypertarget{#1}{}\xspace%
    }%
}

\addto\extrasenglish{%

}

\makeatletter
\def\namedlabel#1#2{\begingroup
    \def\@currentlabel{#2}%
    \phantomsection\label{#1}\endgroup
}
\makeatother

\def\clap#1{\hbox to 0pt{\hss#1\hss}}

\createacronym{alf}{ALF}{Application Library Framework}
\createacronym{alqu}{ALQU}{Algorithms for Quantum Computer Development in Hardware-Software Codesign}
\createacronym{api}{API}{application programming interface}
\createacronym{applib}{AppLib}{Application Library}
\createacronym{bmftr}{BMFTR}{German Federal Ministry of Research, Technology and Space}
\createacronym{clique}{CLIQUE}{Classical Integration of Quantum Computers}
\createacronym{cicd}{CI/CD}{continuous integration and continuous delivery}
\createacronym{dag}{DAG}{directed acyclic graph}
\createacronym{dlr}{DLR}{German Aerospace Center}
\createacronym{dsl}{DSL}{domain-specific language}
\createacronym{efk}{EFK}{Fluentd-Elasticsearch-Kibana}
\createacronym{gui}{GUI}{Graphical User Interface}
\createacronym{hpc}{HPC}{high-performance computing}
\createacronym{ip}{IP}{intellectual property}
\createacronym{mlir}{MLIR}{multi-level intermediate representation}
\createacronym{mqss}{MQSS}{Munich Quantum Software Stack}
\createacronym{nisq}{NISQ}{noisy intermediate-scale quantum}
\createacronym{orm}{ORM}{object-relational mapping}
\createacronym{ornl}{ORNL}{Oak Ridge National Laboratory}
\createacronym{openqasm}{OpenQASM}{Open Quantum Assembly Language}
\createacronym{pite}{PITE}{probabilistic imaginary time evolution}
\createacronym{qaoa}{QAOA}{quantum approximate optimization algorithm}
\createacronym{qci}{QCI}{DLR Quantum Computing Initiative}
\createacronym{qc}{QC}{quantum computing}
\createacronym{qec}{QEC}{quantum error correction}
\createacronym{qem}{QEM}{quantum error mitigation}
\createacronym{qft}{QFT}{quantum Fourier transform}
\createacronym{qir}{QIR}{quantum intermediate representation}
\createacronym{qpe}{QPE}{quantum phase estimation}
\createacronym{qpu}{QPU}{quantum processing unit}
\createacronym{qse}{QSE}{quantum software engineering}
\createacronym{sc}{SC}{Institute of Software Technology}
\createacronym{sdk}{SDK}{Software Development Kit}
\createacronym{se}{SE}{software engineering}
\createacronym{vqe}{VQE}{variational quantum eigensolver}

\newcommand{\QCIConnect}{\tool{QCI~Connect}}
\newcommand{\quark}{\package{quark}}
\newcommand{\quapps}{\package{quapps}}

\title{QCI~Connect: A Modular Full-Stack Quantum Computing Platform}

\author[1]{Eric Bertok~\orcidlink{0000-0003-0753-848X}}
\author[1]{Hannes Busche}
\author[3]{Florian Drinkler~\orcidlink{0009-0005-1463-6653}}
\author[2]{David Emmanuel-Costa~\orcidlink{0000-0002-5068-7125}}
\author[1]{Daniel Herr}
\author[1]{Steffen Hien~\orcidlink{0000-0001-6723-212X}}
\author[2]{Thomas Keitzl~\orcidlink{0000-0003-1552-4766}}
\author[2]{Elisabeth Lobe~\orcidlink{0000-0002-3473-8906}}
\author[3, 4]{Alexandru Paler}
\author[2]{Johannes Renkl}
\author[1]{Martin Rymarz~\orcidlink{0000-0002-8253-2080}}
\author[2]{Gary Schmiedinghoff~\orcidlink{0000-0003-2259-7365}}
\author[2]{Peter K. Schuhmacher~\orcidlink{0000-0003-1232-4363}}
\author[2]{Thomas Stehle}
\author[3]{Benedikt Strobel}
\author[3]{Hanna Tschakert}
\author[3]{Adrian Vetter}
\author[2]{Andre Waschk}
\author[2]{Alexander Weinert~\orcidlink{0000-0001-8143-246X}}
\author[2]{Lukas Windgätter~\orcidlink{0000-0002-4986-8789}}
\affil[1]{d-fine GmbH, An der Hauptwache 7, 60313 Frankfurt, Germany}
\affil[2]{German Aerospace Center~(DLR), Institute of Software Technology, Linder Höhe, 51147 Cologne, Germany}
\affil[3]{PlanQC GmbH, Münchener Str. 34, 85748 Garching}
\affil[4]{Aalto University, 02150 Espoo, Finland}

\date{\normalsize\today}

\begin{document}

    \maketitle

    \abstract{
        In a world of various competing quantum computing architectures,
hardware-agnostic, full-stack platforms are necessary to bring
the full power of quantum computing hardware
to domain experts via the cloud.
\textit{\QCIConnect} and its Software Development Kit provide a reference architecture for a full-stack platform with a modular design and open-source interface definitions,
built to facilitate a community-driven application ecosystem.
Here, we present its overall design and features, central interfaces, and lessons learned, both for users of the platform and as a reference guide for future developments. 
 \\ \\
    }


    \section{Introduction}
\label{sec:intro}

\Gls{qc} has advanced rapidly in recent years,
evolving from the early \NISQ era toward increasingly fault-tolerant architectures
that support mid-circuit measurements, real-time feedback, and elementary error-correction procedures \cite{bluvstein2026faulttolerant, awschalom2025challenges}. 
Alongside the rapid advancement of hardware platforms, numerous institutions and industrial players have developed their own quantum software stacks
~\cite{ibm2026quantumwebpage, amazon2026braketwebpage, microsoft2026azurewebpage, elsharkawy2025integration, burgholzer2026munich, shehata2026bridging}.
Domestically, the research landscape is heavily fragmented -- many isolated, uncoordinated projects whose software components are rarely standardised or interoperable.

The \BMFTR has recognised the importance to harmonise and standardise the efforts within the growing German quantum-computing  landscape.
The final report of the explorative phase of \BMFTR's QC Next~\cite{kunst2025schlussberichtreport} proposes a shared, vendor-neutral reference architecture for the quantum software stack.
It should use standardised open interfaces between its layers, so that components may be exchanged or extended without disturbing the full stack, guarding against the vendor lock-in that entrenches fragmentation. 
They suggest that reference implementations and application-specific blueprints make the architecture concrete, and an active community should be organised around such a stack to carry it toward a de-facto standard. Such consolidation can shorten the path from laboratory to application and strengthen European technological sovereignty. 
The \DLR takes up this challenge within its \QCI, joining forces with the companies \mbox{d-fine} and \mbox{PlanQC} to build such a stack.

\begin{wrapfigure}{r}{0.4\textwidth}
    \vspace{-1.5\baselineskip}
    \include{figures/QCI_Connect}%
    \vspace{-3\baselineskip}
\end{wrapfigure}

\begin{figure}[t]
    \centering
    \def\svgwidth{\linewidth}
    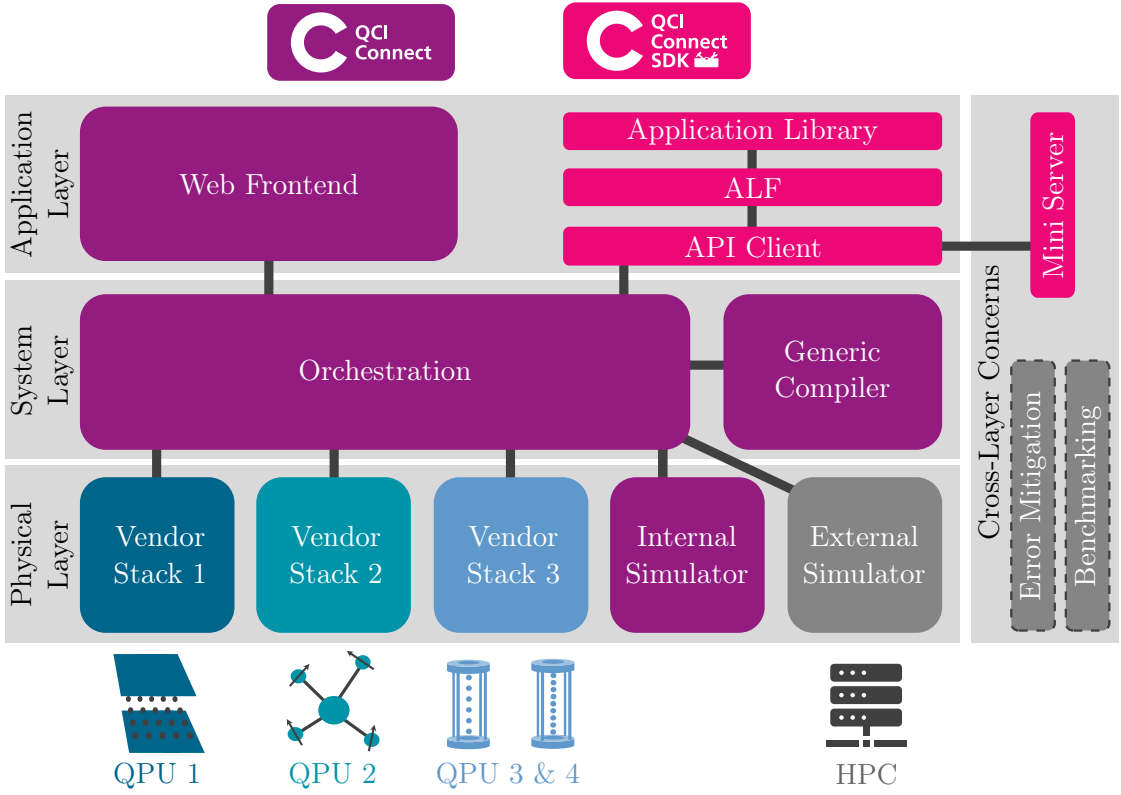
    \caption{High-level overview of \QCIConnect. The software stack serves as a quantum computing gateway for users from academia and industry, within and outside DLR. Major challenges lie in the simultaneous orchestration, scheduling, compilation, and data management for a wide array of simulators and hardware architectures with vendor-specific stacks. The four pink components in the top-right corner form \QCIConnect SDK and are described in more detail in~\autoref{sec:sdk}. The cross-layer concerns of error mitigation communicate between all layers. Benchmarking will be addressed in the future.}
    \label{fig:overview}
\end{figure}

Here, we present a quantum-computing software stack in alignment  with~\cite{kunst2025schlussberichtreport}: \QCIConnect, 
    a modular quantum computing platform with a layered structure, open interfaces and an open-source Software Development Kit.
It serves as a unified entry point for users ranging from academic researchers to industrial end users.
Its layered structure separates the concerns of the full platform stack into a physical layer, a system layer, and an application layer, each introduced in the following paragraphs. 
The separation of concerns together with open interfaces allows the integration of heterogeneous quantum hardware. The high-level perspective is sketched in \autoref{fig:overview}.

The physical layer consists of various vendor stacks, offering a diverse range of hardware architectures including ion traps, neutral atoms, nitrogen-vacancy centers and photonic systems.
The vendors stacks comprise their \QPU hardware with their simulators, specialised control, and compilation software.
To improve the vendor's experience when connecting their stacks, \QCIConnect provides a reference implementation of the backend \API server that can be built upon \autocite{dlr2026qciconnectapi}.

The system layer of \QCIConnect offers an abstraction for submitting and monitoring quantum applications through a hardware-agnostic frontend \API \autocite{dlr2026qciconnectapi}.
Its central component, the orchestrator, tracks quantum jobs as multi-stage workflows and schedules them centrally: 
it coordinates job queues across \QPUs and simulators, enforces scheduling policies, and supports prioritization and fair resource sharing.
A modular integration model encapsulates \QPUs, compilers, simulators, and \HPC-based services as independent backend connectors, so new components integrate with minimal effort.
Through the priorly mentioned unified backend \API definition, the vendor-specific complexity is kept out of the core platform.

The application layer of \QCIConnect builds on this abstraction to deliver the unified entry point for users.
It exposes both a web frontend and a \SDK~\cite{dlr2026qciconnectsdk}, which both build on the frontend \API.
They provide interactive use as well as programmatic access in application workflows.
The \SDK adds a high-level \ALF:
The provided classes simplify the application development to the core logic.
An inheritance-based type system for problems, algorithms, and results increases re-usability and lets users easily search for algorithms that suit a given use case.
The \ALF supplies wrappers that ease the integration of packages based on other frameworks and which we use to build applications that connect to existing \DLR quantum computing libraries.
On request, \ALF applications may be integrated into the \QCIConnect web frontend.

Spanning these three layers lie cross-layer issues spanning from local development, platform security to benchmarking and error mitigation. 
To facilitate local development and testing a so-called mini server provides a sparse implementation of the system layer.
It can be used to run compilers and simulators locally on the user's computer for prototyping before deploying them on the platform. 
In future iterations of \QCIConnect, interfaces should be implemented between generic error mitigation schemes on the application layer and vendor-specific mitigation implementations on the physical layer to allow for better interoperability.
In addition, a full-stack benchmark database can allow profiling and targeted improvements of the platform.

The remainder of this paper is organised as follows.
\autoref{sec:background} provides the institutional and technical background of the \QCI and outlines the \DLR's IT landscape in which \QCIConnect is deployed. It embeds this into the larger quantum computing ecosystem. This context motivates key architectural requirements and constraints addressed in the subsequent sections.
\autoref{sec:architecture} describes the architecture of \QCIConnect in detail. It introduces the system-layer
components, their communication patterns, and the orchestration, job management, and scheduling mechanisms. In addition, this section discusses the supported programming languages and the underlying technology stack that enables scalable and extensible operation.
\autoref{sec:interfaces} focuses on the
interfaces that connect the system-level orchestrator to external components in all layers. In particular, it presents insights into the standardised open-source interfaces for integrating \QPUs and compiler backends, which form the basis for hardware-agnostic execution and expandability.
The graphical user-facing aspects of the platform are described in \autoref{sec:frontend}. This section introduces the \GUI and illustrates typical workflows for quantum job creation, submission, monitoring, and result inspection.
\autoref{sec:sdk} presents the \SDK, which provides programmatic access to \QCIConnect and supports the development of applications, compilers, and simulators to build a community-driven ecosystem. 
The section gives insights into the structure of the \SDK, how to implement applications with it, and how they can be made available to the wider \QCIConnect community.
Finally, \autoref{sec:outlook} outlines future directions and discusses potential extensions of \QCIConnect in the context of evolving quantum hardware, software standards, and hybrid quantum--classical workflows.

    \section{Background}
\label{sec:background}

In the mature landscape of classical computing, software stacks let user describe their algorithms independently of both the data being processed and the underlying hardware.
The software stacks then transform these high-level descriptions into low-level, hardware-specific instructions without requiring further intervention by the user.

In quantum computing, by contrast, software standardization is scarce. The few tools that address it are usually bound to a single layer of a software stack or a single hardware architecture.
An example are standardised \QIRs, which primarily address the problem of expressing quantum circuits in a hardware-agnostic manner. However, they do not provide a holistic solution for hardware access, job scheduling, or execution management. Critical responsibilities, such as queuing, resource arbitration, authentication, and device availability management, are typically handled in an ad-hoc, hardware-specific fashion. As a result, the software stack becomes tightly coupled to the underlying execution environment, limiting scalability and reproducibility.

In this section, we detail the environment in which the \QCIConnect stack operates in three aspects.
First, we give an overview over the \QCI ecosystem in \autoref{sec:background:qci}. Second, in \autoref{sec:background:quantum-computers}, we highlight the different types of quantum computers procured by \DLR. Finally, in \autoref{sec:intro:existing-platforms}, we review existing quantum computing platforms, which \QCIConnect competes with.

\subsection{DLR Quantum Computing Initiative}
\label{sec:background:qci}

To accelerate effective technology transfer and the development of market-ready solutions, in 2022, the Federal Ministry for Economic Affairs and Climate Action (BMWK) has allocated 740 million euro to \DLR over a four-year period for the development of quantum computing technologies in Germany. 
At the heart of this effort is the \QCI, which serves as a platform for bringing together research institutions, start-ups, established industrial partners, and technology developers. 
Approximately 80 percent of the funding is invested directly in industry over a broad range of projects lead by the \QCI~\cite{dlrqci2026projectswebpage}.
By fostering collaboration across the entire value chain, the \QCI is not only advancing the development of quantum computers but also establishing the industrial capabilities, innovation networks, and economic framework required for a sustainable German quantum computing ecosystem. 
This ecosystem-building approach is a defining feature of the \QCI and a key enabler of Germany’s long-term competitiveness in quantum technologies. 

A central ambition of the \QCI is the creation of a unified software stack that gives a wide range of users access to all quantum computers procured by the initiative within 17 hardware projects -- 
not only within \DLR but, in the future, across research, politics and industry.
The stack shall facilitate the seamless submission, execution and monitoring of experiments on any available quantum machine.
This demands for a programmatic interface shared across all heterogeneous quantum hardware, 
where aspects such as job orchestration and scheduling are handled in the background.
In practice, however, the procured \QPUs differ significantly in their accessibility: some expose a vendor-specific \API, while others currently support only manual, on-site operation.
Furthermore, the \APIs that do exist are not standardised across vendors, making unified access non-trivial.
Defining and implementing a common backend \API that abstracts over these differences is therefore a core task of \QCIConnect.

At the same time the platform must integrate with \DLR's existing infrastructure, specifically its existing \HPC systems and identity providers. 
The \HPC resources are required to run quantum circuit simulators, to perform compilation or circuit compression, 
and for the support of future hybrid quantum-classical workflows.
With the geographic distribution of users, quantum hardware, and classical \HPC resources, the infrastructure requirements for \QCIConnect are further shaped in several ways.
As an example, user access management across all sites is handled through \DLR's institutional identity provider, which \QCIConnect integrates with to authenticate and authorise users.

Another goal of this software stack, gaining increasing importance with the advance of the \QCI, is also enabling interoperability, knowledge exchange, and coordinated innovation across the interdisciplinary contributors of the full quantum computing ecosystem with about 60 use-case and software projects and over 100 partners.
However, the diversity of technologies and stakeholders within this collaborative framework thus creates even more demanding requirements on such a software stack~\cite{basermann2024quantum,carbonelli2024challenges}.
The software platform must not only support the integration of heterogeneous quantum hardware, but also diverse software components from all levels of the stack ranging from low-level hardware control, compilers, and algorithms to the real-world applications from industry and within the \DLR in particular.
For this we need to provide development tools and integration mechanisms with well-defined abstraction layers and standardized interfaces that allow contributions from a wide range of partners.
This is what we aim at with the \QCIConnect \SDK and its accompanying application libraries. 

\subsection{QCI Quantum-Computing Hardware Architectures}
\label{sec:background:quantum-computers}

Available hardware architectures differ widely in their qubit connectivity, gate fidelity, coherence times, and scalability, and it remains unclear which of them will ultimately prove most suitable at scale.
To hedge against this uncertainty and gain broad practical experience, \QCI has procured quantum computers built on four distinct hardware architectures~\cite{dlrqci2026technologieswebpage}:

\begin{description}
    \item[Trapped ions] encode qubits in the electronic states of individual ions held in electromagnetic traps.
    They offer high gate fidelity and all-to-all qubit connectivity, but tend to have slower gate operation times compared to other technologies.

    \item[Neutral atoms] use individual atoms held in optical tweezers or optical lattices as qubits.
    This approach offers flexible qubit connectivity through re-configurable trap geometries and is considered a promising path toward large qubit counts.

    \item[Nitrogen-vacancy (NV) centers] exploit defects in diamond crystal lattices as qubits.
    They are notable for their ability to operate at or near room temperature and their potential for quantum networking applications, but face challenges in scaling the amount of qubits.

    \item[Photonic systems] encode quantum information in photons.
    They are well-suited for quantum communication and certain measurement-based computation schemes, but implementing deterministic two-qubit gates remains an open engineering challenge.

\end{description}

In addition to these quantum devices, \QCI also operates multiple classical compute backends, such as analog computers capable of simulating quantum system dynamics through controlled Hamiltonian evolution and digital twins of the above quantum computers.
While these devices are not quantum computers in the strict sense, they offer a complementary approach to studying quantum phenomena and require connection to \HPC systems for their execution.

The wide range of hardware platforms procured by \QCI from different vendors entails a wide variety of native user interfaces.
Hence, \QCIConnect aims to unify access to most \QC hardware and simulated quantum devices procured by \QCI.

\subsection{Existing Software Platforms}
\label{sec:intro:existing-platforms}

A variety of software platforms and tool chains have been proposed to support the provisioning and use of \QC resources, including \QPUs, simulators, compilers, and auxiliary classical infrastructure~\cite{elsharkawy2025integration}. While these platforms pursue a common high-level objective of enabling users to formulate and execute quantum workloads, they differ substantially in architectural scope, deployment assumptions, and operational models.

The \MQSS~\cite{burgholzer2026munich, mqss2026repo} proposes a reference architecture for quantum software systems and identifies key components such as programming environments, compilers, and execution backends. However, \MQSS is explicitly designed around the assumption that quantum computers are used as accelerators tightly coupled to \HPC systems. As a consequence, it assumes centralised governance, co-located hardware and software, and deployment under a single administrative authority. While \MQSS follows a collaborative development model, it does not address cross-location deployment of hardware, interoperability, or the coordinated operation of independently managed installations.

A similar design philosophy underlies the software stack developed at \ORNL~\cite{shehata2026bridging}, which also focuses on integrating quantum devices as accelerators within \HPC environments. Instead of providing an end-to-end platform for managing quantum workloads, the \ORNL approach exposes \APIs that allow applications to explicitly control job submission and execution. The system has primarily been demonstrated in prototype form, and there is limited support for replication, lifecycle management, or operation across multiple locations.

Other widely used tools address specific layers of the quantum software stack rather than full-stack integration. Qiskit \cite{aleksandrowicz2019qiskitsoftware, ibmquantum2026qiskitsoftware, javadiabhari2024quantumpreprint} is a comprehensive \QC library that supports circuit specification, transpilation, compilation, mapping, and routing \cite{ibm2026introductionwebpage}, as well as low-level control abstractions such as fractional gates \cite{ibm2026fractionalwebpage}. While Qiskit is extensively used by developers, it is tightly coupled to IBM’s cloud-based quantum services. The publicly available components primarily target client-side development, whereas the infrastructure required to deploy and operate a comparable service is not openly accessible. As a result, Qiskit is not designed to support independently operated, interoperable \QC platforms across institutions.

Similarly, CUDA-Q \cite{kim2023cuda} provides an embedded \DSL in C++ and Python for quantum program specification, along with an \MLIR-based compiler infrastructure \cite{lattner2021mlir}. Its primary emphasis lies on circuit construction, compilation, and simulation, particularly targeting GPU-based backends. Although execution on selected cloud-based \QPUs is supported, aspects such as resource federation, orchestration, and lifecycle management are outside the main scope of the platform. OpenQL \cite{khammassi2021openql} follows a comparable approach by offering a \DSL and compiler toolchain for quantum circuit specification and optimization, without addressing execution management or platform interoperability.

There also exists work towards a reference architecture for \QC software stacks developed as part of the projects of QC Next~\cite{kunst2025schlussberichtreport}.
This reference architecture comprises an application layer, a system layer, and a physical layer, each of which in turn comprises multiple software building blocks that translate quantum algorithms for execution on physical hardware.
At the time of writing, implementation of individual building blocks is underway as part of the QC Next main project named ``FullStaQD'', while satellite projects aim to validate this proposed architecture via partial implementation in industrial use cases~\cite{fraunhoferiao2026fullstaqdwebpage}.
Within this framework, the high-level programming language Eclipse Qrisp~\cite{seidel2024qrisppreprint, fraunhoferfokus2026eclipsesoftware} shall also be advanced and integrated in the reference architecture. 

On a less technical level, there exists recent work by the German Institute for Standardization to standardise the interfaces between quantum computer backends and software frameworks~\cite{2025dinstandard}.
This work, however does not comprise reference implementations nor technical references, but instead only specifies the content to be passed via this interface.
Thus, merely adhering to the standard does not ensure technical compatibility of the communicating systems.

In summary, existing \QC platforms, software stacks and standardization efforts predominantly focus on isolated components of the quantum software ecosystem or assume operation within a single administrative domain. Unified access to heterogeneous resources, and systematic orchestration remain largely unaddressed. These limitations motivated the development of the platform \QCIConnect, which seeks to reduce fragmentation and simplify access to quantum computing infrastructures.

    \section{Architecture}
\label{sec:architecture}

This chapter covers the technical architecture of \QCIConnect, building on the basic \DLR IT infrastructure described in \autoref{sec:background}, which motivates and explains parts of the chosen architecture.

\begin{figure}[p]
    \centering
    \def\svgwidth{0.93\linewidth}
    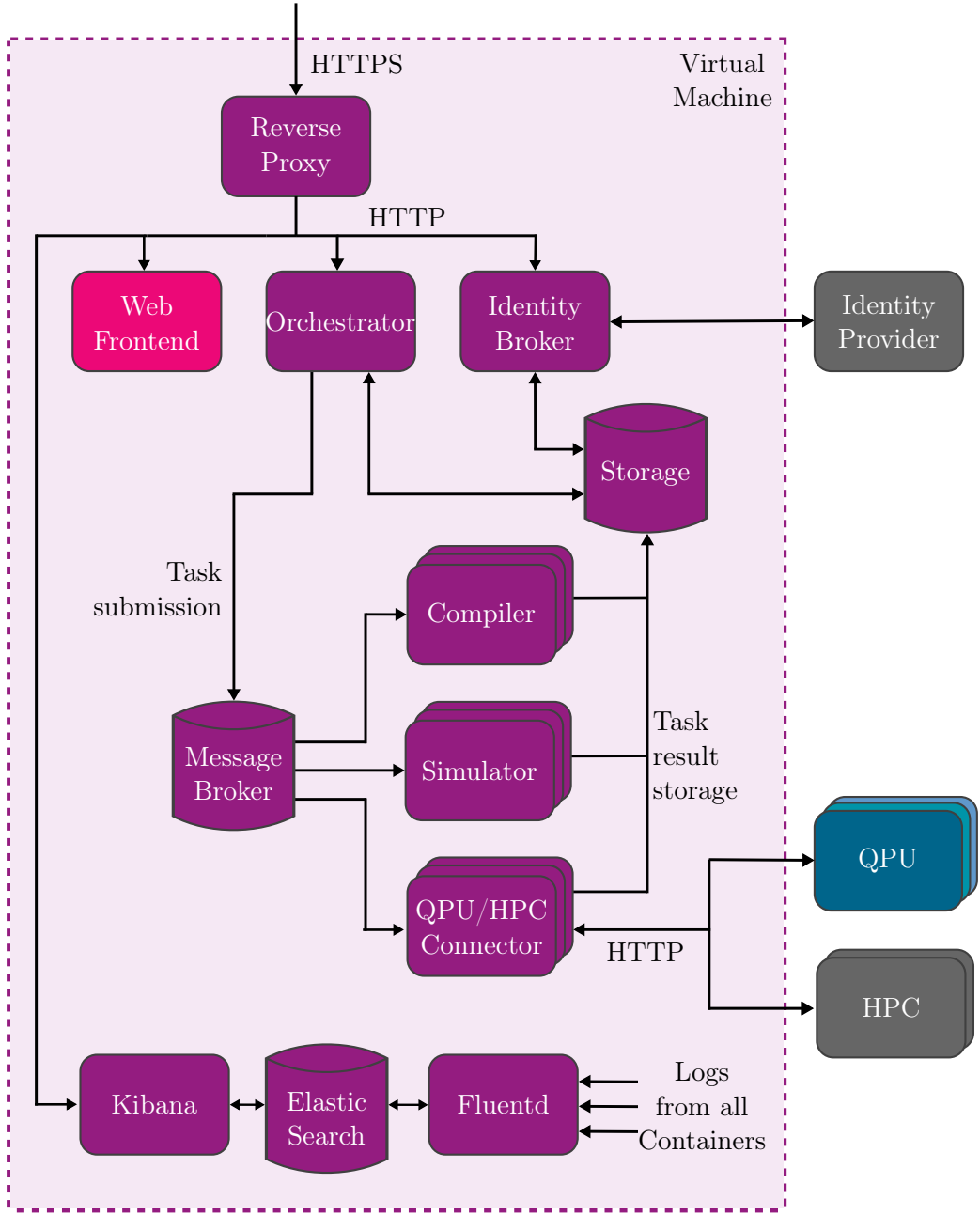
    \caption{High-level architecture of \QCIConnect, showing the containerised service stack hosted on a dedicated virtual machine inside the DLR network. User requests are routed through a reverse proxy to the frontend, orchestrator, and identity broker. The orchestrator manages asynchronous task execution through a message broker and dedicated worker services for compilation, simulation, and QPU or HPC access, with persistent storage, centralised logging, and integrations with external identity, QPU, and HPC systems.}
    \label{fig:architecture_diagram}
\end{figure}

The high-level architecture diagram, \autoref{fig:architecture_diagram}, illustrates the main components of the platform, their interconnections, and integrations with external services: the \DLR user management system (Shibboleth), the connected \QPUs, and the \HPC systems. \QCIConnect is hosted on a dedicated virtual machine inside the \DLR network, and it is implemented as a modern, service-oriented application that is fully containerised using Docker. 

\subsection{Components and Communication}

A key design goal of \QCIConnect is the flexibility to integrate additional compilers, \QPUs or \HPC systems. Therefore, the component architecture is deliberately modular: each compiler or \QPU (including simulators and \HPC systems) is accessed via a dedicated connector that is encapsulated as an independent service and registered with the orchestrator through a configuration interface. This structure ensures a clear separation of concerns, simplifies testing and maintenance and isolates potential faults or performance issues of a single backend from the rest of the platform. As long as a new component adheres to the agreed system interfaces (cf. \autoref{sec:interfaces}) and task schema, it can be added, removed, or replaced without modifying the rest of the platform. 

Incoming user traffic is routed through a reverse proxy, which serves as the only externally exposed entry point to the platform. Graphical user access is provided through an edge layer composed of this reverse proxy and a web server. An nginx-based reverse proxy terminates TLS connections, enforces fundamental security policies, and routes incoming HTTP(S) traffic to the appropriate backend services. Static assets of the web frontend are delivered by a separate nginx instance optimised for content delivery. The frontend itself is implemented as a single-page application that communicates with the backend via a well-defined RESTful \API; see~\autoref{sec:frontend}.

The core backend of \QCIConnect consists of the orchestrator service and a set of specialised worker services. Exposed through the reverse proxy, the orchestrator provides the main public \APIs used both by the graphical user interface and by the \QCIConnect \SDK introduced in~\autoref{sec:sdk}. At the heart of the system, the orchestrator collects user requests, validates and stores them, and translates high-level job descriptions into sequences of executable tasks. These tasks are submitted to a message broker that provides asynchronous, decoupled communication between the orchestrator and the worker services and forms the backbone of the internal job-management infrastructure. Worker processes implemented using Celery pick up tasks from the broker. Dedicated worker containers handle specific types of tasks, such as:
\begin{itemize}
    \item compilation of high-level quantum programs or circuits into the target representation required by a specific \QPU,
    \item classical simulation of quantum circuits for testing, benchmarking, and debugging,
    \item interaction with DLR \QPUs, and
    \item interaction with DLR \HPC systems that provide resource‑intensive simulators and digital twins of (future) \QPUs.
\end{itemize}
This separation into focused services improves modularity and facilitates the integration of new technologies.

The internal communication within \QCIConnect follows an event-driven pattern based on a central message broker that is implemented using Redis. Application services such as the orchestrator submit tasks to Redis, which are stored in queues and consumed by Celery worker processes running in one or more worker containers. Because task submission and execution are decoupled, the orchestrator remains responsive even when individual tasks are long-running or when temporary overload occurs. The number of worker processes can be scaled up or down according to the demand, enabling horizontal scalability. Failed tasks can be retried in a controlled manner, and message persistence ensures that no work is lost if a worker or container has to be restarted.

Persistent state and metadata are stored in a relational database that is currently implemented using PostgreSQL. Typical data include user and project information, job specifications, workflow graphs, execution logs, and the results of successfully completed tasks. The message broker queues tasks and coordinates the work of Celery workers. 

Identity and access management are handled by an identity broker (Keycloak) that is integrated with the \DLR-wide identity provider (Shibboleth). Authentication is delegated to the institutional identity provider, which issues assertions about the users' identities and organisational affiliations. The identity broker translates these assertions into authentication tokens and platform-specific roles, which are then used by the orchestrator to authorise access to projects, resources, and \QPU backends.

To ensure observability and operability, the platform includes a centralised logging and monitoring stack. Application and infrastructure logs are collected by Fluentd, stored in Elasticsearch, and visualised via Kibana. This stack, known as \EFK stack, enables developers and operators to inspect system behaviour, diagnose issues, and audit all containers and services uniformly.

For internal tracking and control, the platform collects and aggregates various usage statistics, including user activity, job characteristics, and \QPU utilisation. These metrics are processed exclusively for monitoring, capacity planning, and reporting, and do not affect the behaviour of individual jobs. Similar to the \EFK-stack, access to such statistics is strictly limited to authorised administrative users and is not available to standard platform users.

\subsection{Orchestration, Job Management, and Scheduling}

From the users' perspective, the interaction with \QCIConnect starts by defining a job through the web frontend or via the \SDK. Alternatively, also the public \API can be used. Conceptually, a job represents a quantum workflow composed of one or more tasks. Such a chain of tasks may include compilation tasks (e.\,g., translating a high-level quantum program into a hardware-specific representation), \QPU execution tasks, classical simulation tasks, and pre-/post-processing tasks. Internally, the orchestrator represents each job as a \DAG of tasks with explicit dependencies.

When a new job is submitted, the orchestrator validates the job specification, checks permissions and resource availability, and then persists the job and its \DAG in the database. For each task whose dependencies are already satisfied, the orchestrator creates a task message and enqueues it in the appropriate Redis queue. Worker services continuously consume tasks from their respective queues, execute them, and report back status and results to the orchestrator. This procedure allows complex multi-stage workflows to be expressed and executed without manual intervention.

The platform supports flexible task chaining, enabling users to build complex workflows that combine several compilers, multiple \QPUs, and (post-)processing steps. For example, a typical workflow may compile a quantum circuit using a generic compiler, apply hardware-specific optimisation passes, execute the resulting circuit on a chosen \QPU, and finally perform classical aggregation, analysis, and visualisation of the results. The architecture is designed such that additional (post-)processing tasks can be seamlessly inserted into existing compiler--\QPU chains, enabling the implementation of hybrid algorithms and advanced error-mitigation strategies.

Each \QPU and compiler manages incoming tasks through a queuing system that incorporates a priority mechanism, ensuring fair access to the resources across the entire user base. For this purpose, every \QPU or compiler task is assigned a priority flag (low, medium, or high), which is automatically determined based on user privileges as well as the execution history of previous runs, allowing the platform to prevent individual users or projects from monopolising scarce \QPU resources. The scheduler within the orchestrator respects these priorities when selecting the next task for submission, ensuring that high-priority tasks can be processed sooner while still maintaining fairness across the user base. 

Users retain control over their jobs throughout the life cycle. Via the web frontend or \SDK, they can stop running jobs if the \QPU supports this action, cancel jobs that are still queued, and inspect the current status of all their jobs. Each task within a job can be in states such as ``pending'', ``running'', ``succeeded'', ``failed'', or ``cancelled'', and these task-level states are aggregated to provide an intuitive job-level overview. Completed jobs expose their results in a structured format that can be downloaded or directly visualised via the web frontend. For convenience, previously submitted jobs can be easily resubmitted or used as a basis for the input of new, similar jobs.

\subsection{Supported Languages}

A central challenge in building a unified quantum software stack is defining a consistent and interoperable representation for quantum circuits.
These circuits flow through multiple layers of the software stack: from user input and compilation, to execution on quantum hardware or simulators. These layers vary in their requirements: some treat circuits as opaque data (e.g., for storage, transmission, or routing), while others interpret and transform them (e.g., for optimization, mapping, or low-level compilation).

To ensure synergy across these layers, we adopt a canonical representation for quantum circuits within the platform. This representation serves as a common language for internal communication and enables predictable, modular design. After evaluating several options, we selected \OpenQASM~\cite{cross2017openpreprint} as our canonical format.

\OpenQASM is a domain-specific, low-level language for describing quantum circuits. It provides a human-readable textual representation of quantum operations, including gate applications, qubit measurements, and classical control flow. As an open standard, \OpenQASM is widely supported across quantum computing frameworks and tools, making it a natural choice for interchange between components of the stack.

The more recent \OpenQASM~3~\cite{cross2022openqasm} extends the language with advanced classical control features such as real-time classical computation and dynamic circuit constructs. 
\OpenQASM~3 is the default circuit format in \QCIConnect and it is supported by all \QPUs, \HPCs and simulators that are integrated in the platform. Any job can therefore always be submitted in \OpenQASM~3, and if no explicit format is specified in the \API, \OpenQASM~3 is assumed by default.
However, we currently restrict the canonical representation to the gate-set subset of \OpenQASM~3 — a backward-compatible subset that excludes dynamic classical control. This decision is driven by hardware limitations: the quantum hardware currently integrated into our platform does not support execution of classical instructions at the control-electronics level, rendering these features impractical for production use. 
However, this design preserves a clear path toward full \OpenQASM~3 adoption as hardware capabilities evolve.

Beyond the canonical format, \QCIConnect is fully language-agnostic in its user-facing workflows. Each \QPU, simulator, or \HPC-based backend advertises the set of circuit formats and intermediate representations it supports (e.g., \QIR, vendor-specific formats, or ecosystem-specific DSLs). Users are free to choose the language or representation they prefer, provided it is supported by the target \QPU or its upstream compiler.

The chosen format is indicated explicitly in the job submission via a format enumeration, and \QCIConnect forwards the circuit in the requested format without requiring conversion to a single internal representation. Compatibility checks in the orchestration and service layers ensure that only \QPU–-compiler combinations sharing at least one supported format are used. Optional compiler services can be integrated to translate between formats when needed, but such translations are not mandatory.

\subsection{Technology Stack}

The implementation of \QCIConnect relies on a modern and widely adopted technology stack. Containerisation is provided by Docker, which allows each component to run in its own isolated environment and simplifies deployment across different stages (development, testing, production).

All public and internal \APIs are specified using OpenAPI. These specifications serve as the single source of truth for the interface contracts between services and are used to generate client libraries and server stubs via \package{openapi-generator}~\cite{2026openapigeneratorsoftware}. This approach reduces manual coding effort and helps to keep the documentation, implementation, and clients in sync.

The backend services are written in Python. FastAPI is used as the main web framework for implementing high-performance, type-annotated RESTful \APIs. Task execution and distributed job processing are implemented using Celery on top of Redis as the message broker. SQLModel provides an \ORM layer on top of \mbox{PostgreSQL}, combining Pydantic-style data validation with SQLAlchemy-style persistence, while PostgreSQL serves as the primary relational database for storing persistent platform data such as job metadata, user and project information, and configuration state. Together, these libraries enable concise and maintainable backend code.

Within the \QCIConnect architecture, nginx is employed in two roles. On the one hand, it acts as a reverse proxy at the platform's edge, responsible for TLS termination, routing, and basic request filtering. On the other hand, it functions as a web server optimised for serving the static assets of the frontend application. The frontend is implemented using React and TypeScript, providing a modern responsive and interactive user interface that communicates with the backend exclusively through the defined \APIs.

For logging and observability, the platform uses the \EFK stack: Elasticsearch for log storage and indexing, Fluentd for log collection and forwarding from Docker containers, and Kibana for interactive exploration and visualisation. This stack provides a comprehensive view of the platform's behaviour and supports debugging, performance analysis, and compliance requirements.

Finally, third-party quantum software packages such as Qiskit~\cite{ibmquantum2026qiskitsoftware} are integrated into \QCIConnect to provide simulation capabilities and circuit transformation. These packages are encapsulated within dedicated containerised worker services, ensuring that their dependencies and runtime environments do not interfere with other platform components.

Overall, this technology stack combines robustness, performance, and extensibility, positioning \QCIConnect as a flexible platform for current and future quantum computing workflows.
    \section{System Interfaces}
\label{sec:interfaces}

The communication with \QCIConnect relies on several interfaces, which we group into system interfaces and user interfaces. In this section, we describe the system interfaces, which handle communication between the orchestrator (\eg, via dedicated connectors; \cf \autoref{sec:architecture}) and other technical systems, in particular, QPUs and compilers. The user interfaces, on the other hand, consist of a \GUI and a \SDK, and are introduced in \autoref{sec:frontend} and \autoref{sec:sdk}, respectively.

As described in \autoref{sec:architecture}, the design of \QCIConnect provides connectors for the communication with QPUs. In the chosen setting, each connector is dedicated to an individual QPU, and the same RESTful \API is used for each QPU, independent of its details and specifics.\footnote{Within each QPU connector, there remains the flexibility to adapt and customise the underlying API to meet specific requirements and short-term adjustments.} Consequently, real quantum hardware, classical simulators, digital twins, HPC systems, etc., share the same API. The use of such a universal API is only possible due to careful engineering and close alignment with the involved QPU vendors.
The specifications of the \API are publicly available~\cite{dlr2026qciconnectapi} and revised when an update is needed.

In its current version\footnote{The latest API version at the time of submission is v.0.4.0.}, the most important endpoints comprise (but are not limited to):
\begin{itemize}
    \item GET /qpu: Get information about the qpu and its current state.
    \item POST /qpu/jobs: Upload new job to QPU.
    \item GET /qpu/jobs/\{job{\_}id\}/result: Get result of a job that was run on QPU.
\end{itemize}
These three endpoints suffice to retrieve most of the information about the QPU, post a new quantum job, and return its result. Additional endpoints that are not explicitly listed here simplify these steps and considerably improve the performance. For example, the \code{QPU} object retrieved by the /qpu GET endpoint is extensive, and dedicated endpoints exist for getting selective information, e.g., on the status of the QPU only. For this purpose, as indicated by the API specification~\cite{dlr2026qciconnectapi}, the \code{QPU} object itself contains several properties organised as sub-objects, such as \code{info}, \code{gate\_set}, \code{gate\_fidelities}, and \code{qubit\_connectivity}. 
    \section{Graphical User Interface}
\label{sec:frontend}

The \GUI of \QCIConnect is a single-page web-application available at the domain \hyperlink{https://connect.qci.dlr.de/}{https://connect.qci.dlr.de/}. It allows for quick actions such as a live monitoring of \QPU statuses and occupations, and guides the users through the creation, submission, and retrieval of quantum jobs. While more complex and computationally intensive quantum jobs are arguably managed primarily with the \SDK described in \autoref{sec:sdk}, the \GUI is mainly designed for visual demonstrations, trainings, application of use cases, etc. It thus provides the users with a natural introduction to \QCIConnect and enables a seamless interaction with the quantum computers of the \QCI. Even for power users that submit jobs via the \SDK, the \GUI offers a convenient way to organise, filter and manage submitted jobs.

Technically, the web application uses nginx as a web server and it is built with TypeScript, making use of the React library. Both the \GUI and the \SDK use the same RESTful \API to communicate with the orchestrator of \QCIConnect. This allows users to switch seamlessly between intuitive, browser-based access and programmable, script-based workflows.

\subsection{Quantum Job Creation}

\begin{figure}[t!]
    \centering
    \includegraphics[width=1\textwidth]{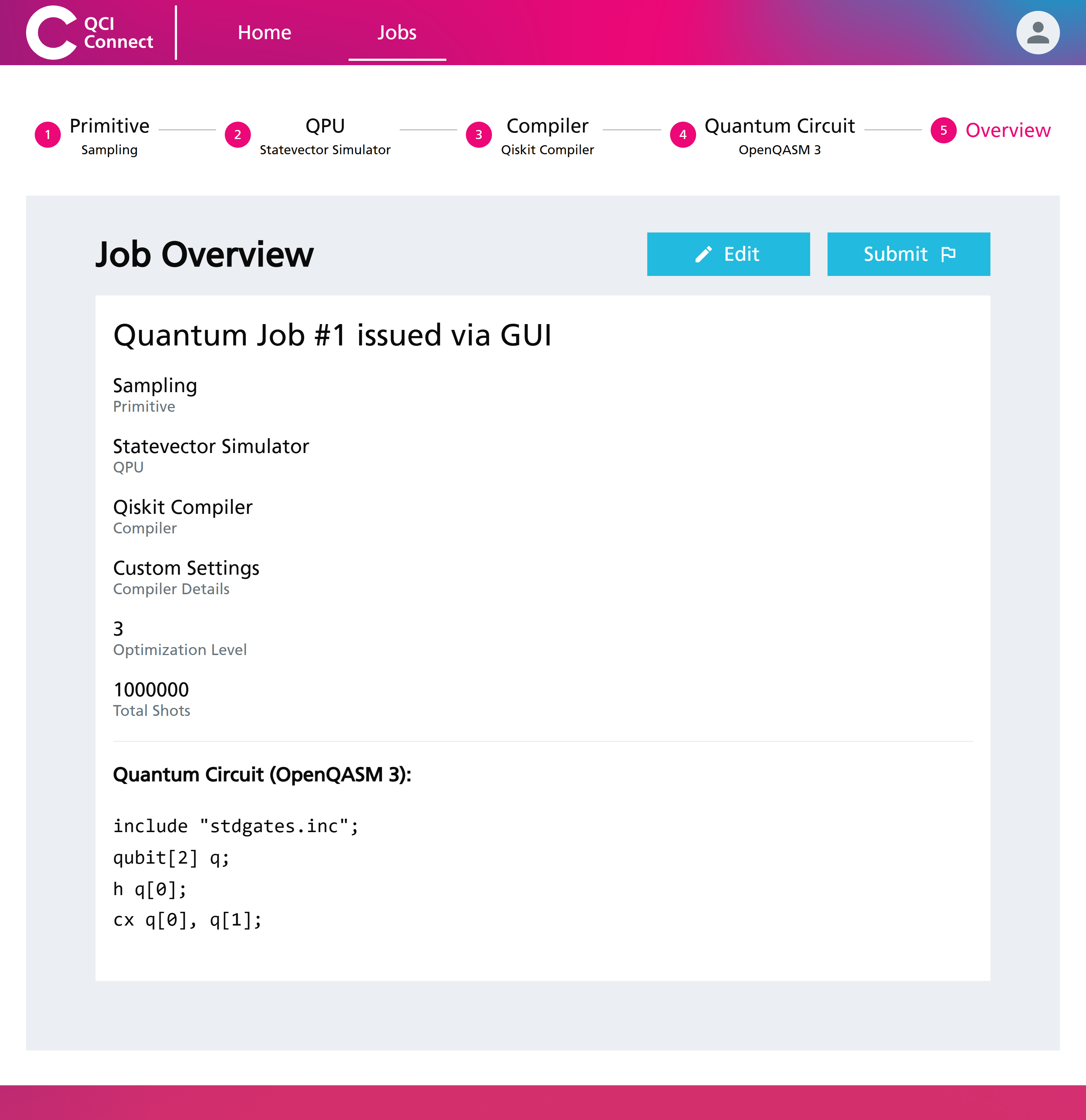}
    \caption{Screenshot of the overview of a quantum job prior to its submission via the GUI of the \QCIConnect web interface.}
    \label{fig:overview_page}
\end{figure}

The creation and submission of a new quantum job follows a linear creation process that consists of five steps. \autoref{fig:overview_page} shows the overview of a previously created quantum job prior to its submission to the selected \QPU.

The linear job creation process consists of:
\begin{enumerate}
    \item Primitive Selection
    \item \QPU Selection
    \item Compiler Selection
    \item Quantum Circuit Definition
    \item Job Overview
\end{enumerate}
This linear process is simple and intuitive, allowing for easy quantum job submissions as part of interactive demonstrations while retaining near feature parity with the \SDK.

The \QPU selection page provides a concise overview of all \QPUs currently integrated into the platform. At a glance, users can see which \QPUs are available to accept new tasks, how many tasks are already queued, and how many qubits each \QPU offers. Expandable cards optionally provide detailed information like accepted and native gate sets, shot limits and qubit connectivities and fidelities.

Compiler selection is based on the supported languages and intermediate representations of a \QPU and includes auto-completion of compiler settings to match the selected \QPU. 

During the successive progression of job creation, validation checks are performed within the \QCIConnect orchestrator prior to proceeding with the next step. For example, if a job is created without a compiler and the instructions on the quantum circuit contain gates that are not supported by the selected \QPU, job validation will trigger an alert and deliver real-time feedback to the user via the GUI. These mid-creation job validations prevent users from submitting ill-defined quantum jobs to \QPUs, and offer guidance for correction, saving not just time but also valuable resources.

\subsection{Quantum Job Management}
\label{sec:frontend_jobmanagement}

\begin{figure}[t!]
    \centering
    \includegraphics[width=1.0\textwidth]{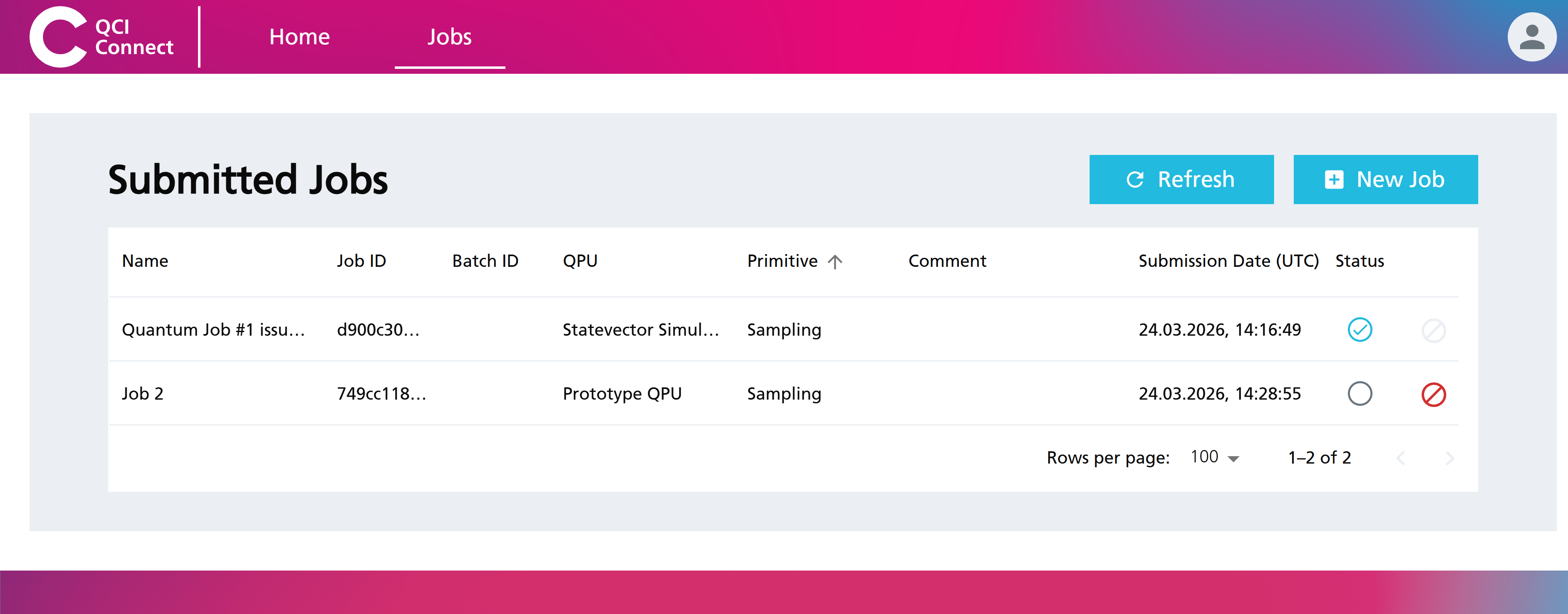}
    \caption{Screenshot of the table of submitted quantum jobs in the \QCIConnect web interface.}
    \label{fig:submitted_jobs}
\end{figure}

The \GUI of \QCIConnect also allows inspecting all previously submitted quantum jobs, providing users with a quick look-up table containing condensed information about their quantum jobs as well as the possibility to monitor the status and to cancel queueing jobs; see~\autoref{fig:submitted_jobs}.

\begin{figure}[p!]
    \centering
    \includegraphics[width=0.89\textwidth]{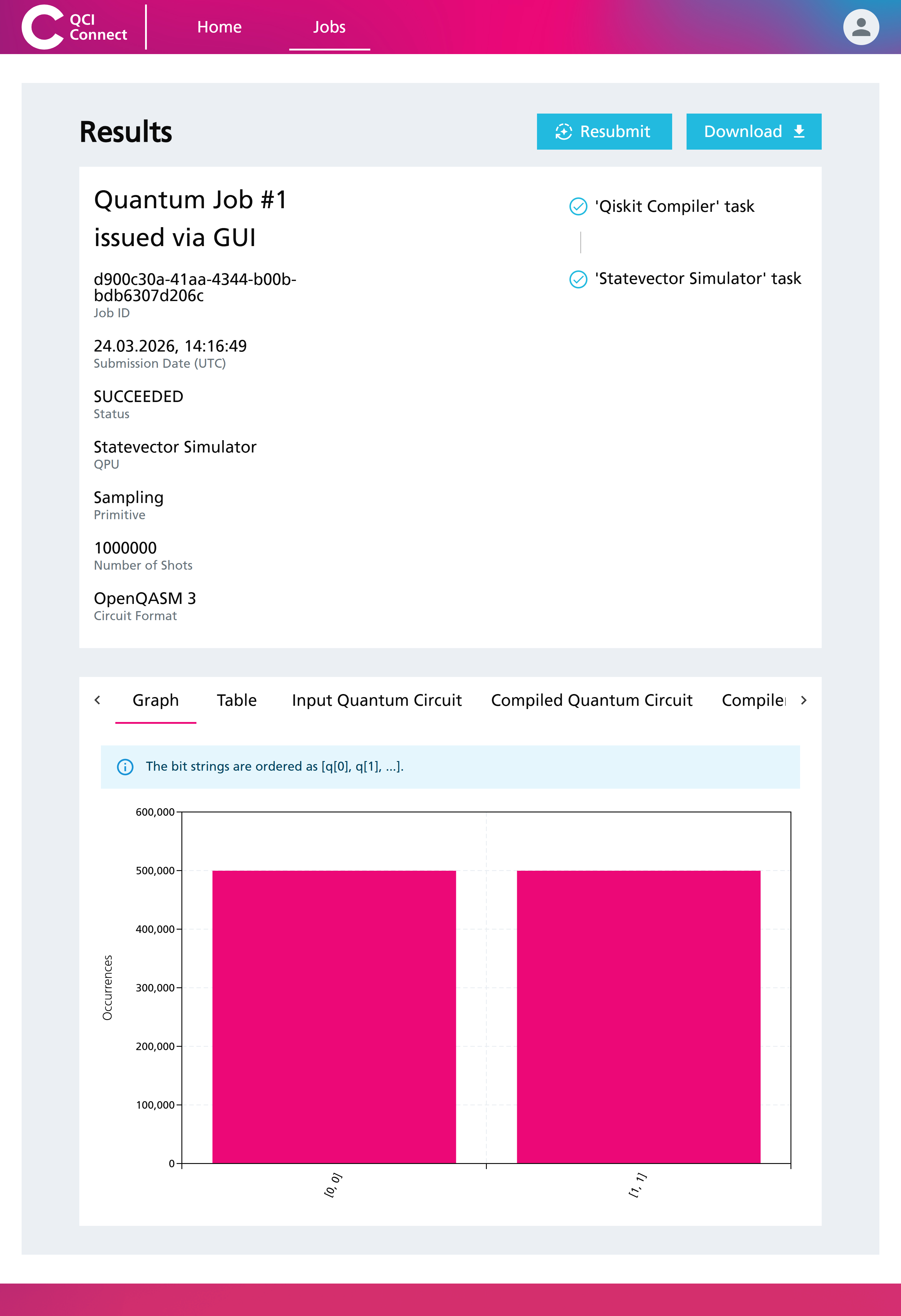}
    \caption{Screenshot of the visual representation of the quantum job results in the \QCIConnect web interface. This page also includes the job specifications and more meta information, \eg, runtime details of the QPU.}
    \label{fig:results_page}
\end{figure}

Selecting a successfully executed quantum job from this list provides the user with information on the results that are graphically available in the GUI and also offered as a download; see \autoref{fig:results_page}. In case of a failed quantum job, the user is informed about the reason of the failure. Moreover, users can quickly resubmit a job based on a previous submission.

    \section{Software Development Kit}
\label{sec:sdk}

\subsection{Overview}

\begin{figure}[b!]
    \centering
    \def\svgwidth{\linewidth}
    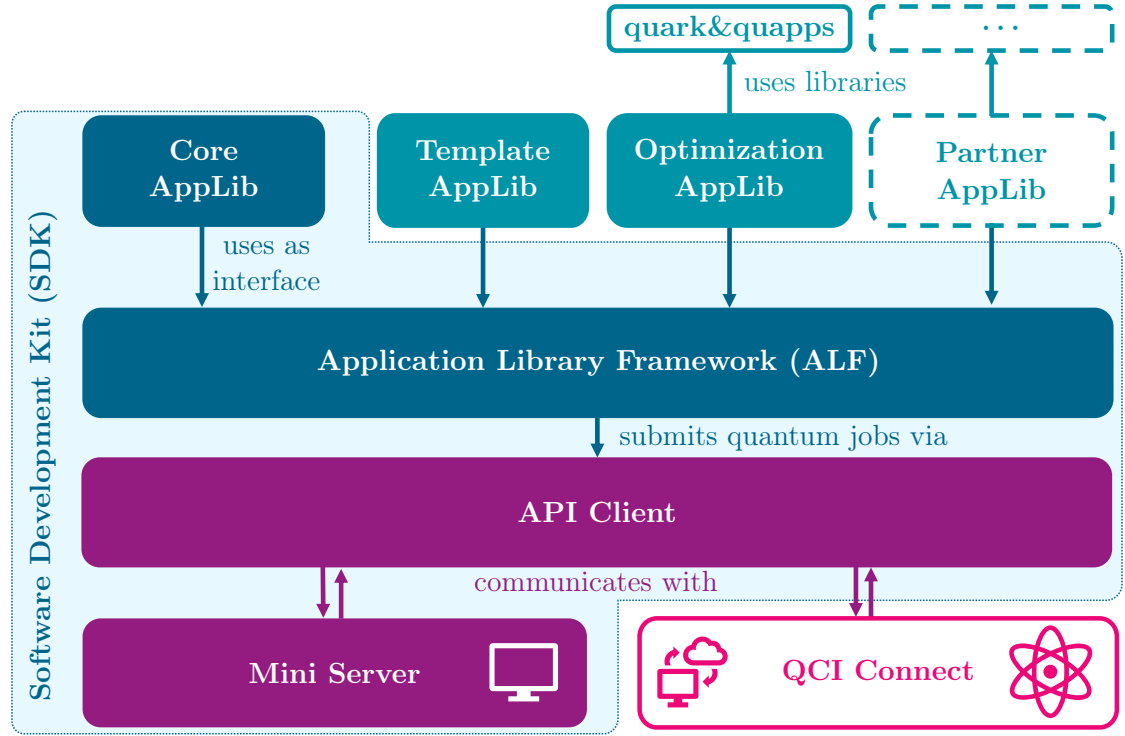
    \caption{SDK architecture for developing IP-protected quantum applications. 
    The API Client communicates with \QCIConnect and the local mini server through the same API. 
    ALF uses this interface to run quantum jobs and provides the interface for applications, while the Core AppLib provides reference implementations. 
    Partners can build extensions to the AppLib (for example starting from the Template AppLib) in their own repository using ALF interfaces and their own existing libraries.
    An example of this is the Optimization AppLib that brings optimization application from the packages \quark and \quapps to \QCIConnect.
    }
    \label{fig:sdk}
\end{figure}

A graphical user interface can only provide limited access to all hardware features and programming capabilities. 
We want \QCIConnect users to address the hardware with as much control and feedback as possible and facilitate a community-driven application ecosystem.
To that end, we provide a \SDK as an open-source Python package under the Apache license~\cite{dlr2026qciconnectsdk} for both experts and non-experts in quantum computing. Its features include:
\begin{itemize}
\item programmatic access to \QCIConnect resources,
\item a hardware-agnostic framework for developing applications and integrating compilers and simulators while protecting \IP through granular access controls,
\item a local execution environment independent of the cloud,
\item reference implementations of standard algorithms and applications in an application library,
\item and tutorials on the usage of all its components,
\end{itemize}
provided by four main components:
\begin{enumerate}
    \item \emph{\API Client}: Low-level interface to \QCIConnect,
    \item \emph{Mini Server}: Local execution environment eliminating cloud dependency during development of compilers or simulators,
    \item \emph{\ALF}: Interfaces for high-level application development,
    \item \emph{\AppLib}: Basic application implementations and curated examples,
\end{enumerate}
\autoref{fig:sdk} illustrates these components
and their interactions. The technical documentation is available via the landing page~\cite{dlrsc2026qciconnectlandingpage}.

Rather than duplicating existing sophisticated quantum libraries, the \SDK aims at easing the integration of existing solutions into the \QCIConnect ecosystem, e.g. , by accepting Qiskit and OpenQASM circuits through a circuit wrapper.
Making the source code available under an open source license is crucial for building a strong community. 
In addition, the extensive documentation and 
the numerous tutorials in the form of Jupyter notebooks simplify the adoption. 
In the following, we describe the individual components in more detail and explain how applications can be built on top of them.

\subsection{API Client \& Mini Server}

The \API Client is a Python library that provides low-level programmatic access to \QCIConnect.
Users can retrieve information on the available \QPUs and compilers, 
  send quantum circuits in OpenQASM format (and in the future QIR) via blocking or non-blocking jobs, and retrieve job results.
The result type depends on the job request: It can contain compiled circuits, measured bit strings, or simulated state vectors.
Jobs from the \API Client and the web frontend share the same backend, see \autoref{sec:architecture},
  ensuring consistent job visibility and monitoring across interfaces.

The mini server is a local backend to the \API Client that communicates with local compilers and simulators. 
This allows the local development and testing of quantum applications, compilers and simulators for the platform.
In addition, this allows users without \QCIConnect authentication tokens to work with the \SDK locally. This is particularly useful for algorithm and code prototyping and allows users to test their implementation before running it on \QCIConnect.

\subsection{ALF \& Core AppLib}

The key for developing a full software stack is modularity that enables specialists in different fields to enhance their specific modules without necessarily needing to know about other modules apart from the interfaces connecting them. For instance, hardware-agnostic quantum algorithm developers in general want to implement their specific new algorithm without having to think about how the circuit is compiled and finally sent to a quantum device. In the ideal case, the algorithm can be reused in multiple different scenarios.  
E.g. by practitioners who want to evaluate quantum computers for their specific application and might just want to choose between different algorithms, error mitigation techniques, compilation features and more and plug them together for a full execution scheme. 
This interoperability can be achieved through suitable abstractions. Those should however not introduce too much overhead with regard to the still limited resources of quantum hardware.

Our package \ALF shall exactly provide such interfaces. It is tailored to the development of applications from the users' perspective and provides a high-level software framework which covers all components from the application to the quantum hardware layer.
Commonly, an application specifies a \emph{problem input},
such as a Hamiltonian and observables, 
and returns \emph{result} objects, which may contain quantities such as expectation values or solutions to optimization problems. 
\ALF automates circuit synthesis, observable measurement (including basis transformations and shot-based expectation-value estimation), communication through the \API Client, and error mitigation. 
It follows a modular structure. 
Through clear and consistent \API definitions across its different layers, \ALF provides components such as error mitigation and circuit construction methods 
which can individually be exchanged without changing the underlying algorithm of the application.
Users can extend the toolbox with their own components, such as mitigation methods on the sampling and estimation level.

\begin{figure}[b!]
    \centering
    \def\svgwidth{\linewidth}
    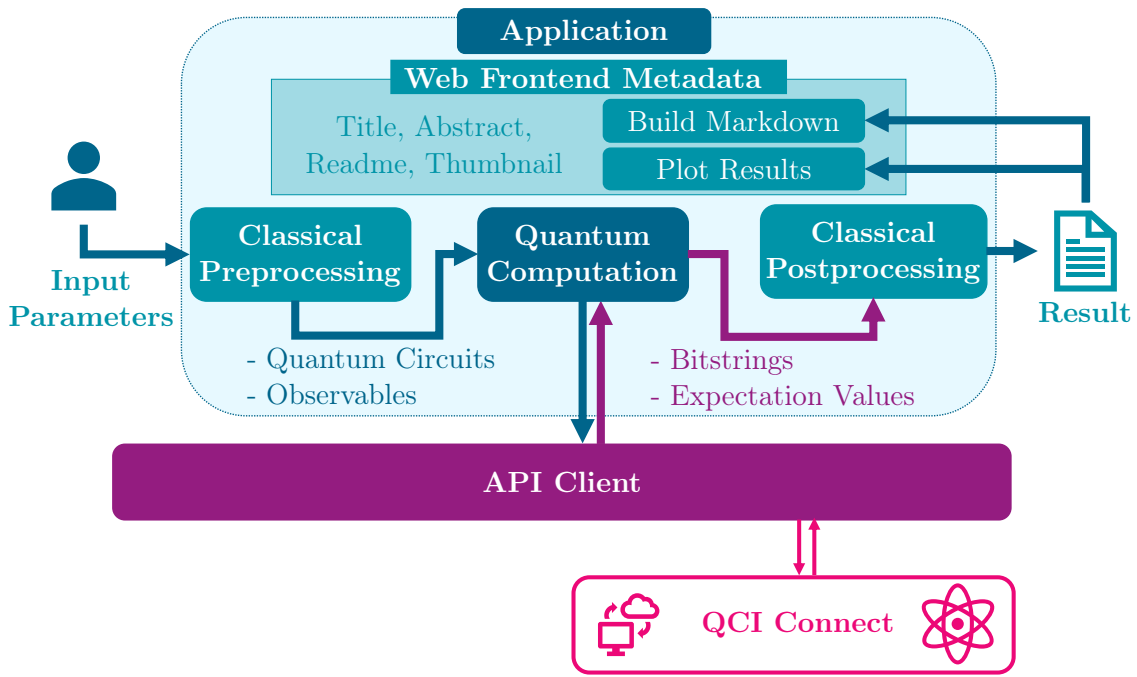
    \caption{Layout of simple classical-quantum-classical workflow of an application. 
    The application developer needs to implement the preprocessing on the input parameters (the problem definition and application parametrization), and the postprocessing of the quantum computation result to the final application results. 
    Further metadata can be defined to enable a web demo in the GUI.}
    \label{fig:alf-app-workflow}
\end{figure}

\autoref{fig:alf-app-workflow} illustrates the classical-quantum-classical workflow for standard applications,
where developers define input parameters, result types, and classical pre- and postprocessing logic.
Beyond these classical-quantum-classical workflow, application developers can 
directly leverage \ALF's tools for sampling and estimating
to implement flexible workflows including out-of-coherence-time hybrid algorithms,
which 
can freely alternate between
classical and quantum computations.

To enable web demonstrations of applications that can be hosted directly on \QCIConnect,
\ALF provides also interfaces for defining metadata that is relevant for the \GUI.
This includes documentation and visualization information about the application,
such as methods for creating figures and Markdown files 
to visualise the results of an application which has been run in the web frontend.
To facilitate these web demos,
an \ALF server is currently being set up that will provide curated applications 
through a communication interface with \QCIConnect.

Envisioning libraries of reusable algorithmic building blocks being fully compatible with each other, 
Core \AppLib provides numerous standard application examples based on \ALF.
This \AppLib will be complemented with several domain-specific application libraries, 
where several use cases from \DLR as well as from the industry partners provide the problem input, as explained in the following section.
This way, the \QCIConnect \AppLib ecosystem can be continuously enhanced with community-driven applications.

\subsection{Workflow for Implementing Application Libraries}
\label{sec:sdk:implement-app}

With such a diverse ecosystem as the QCI and their different partners within and outside of \DLR, a single overarching software library is not suitable and rather multiple domain fields develop their individual packages. If built on top of \ALF, they will however still be able to leverage the full potential of the QCI quantum computers, include their existing individual software packages but keep possible \IPs internal and just share the information they want. 

Partners can extend the Core \AppLib by implementing their own application libraries in a separate repository.
The separation of repositories achieves three critical objectives: 
\begin{enumerate}
    \item \emph{Independent repository management:} Each software project has different requirements on who may access, modify and distribute the source code, and on how the development workflow must be organised. Partners that work on their \AppLib extension are enabled to make these crucial decisions independently from the \SDK repository.
    \item \emph{Minimised dependency conflicts:} Application developers are encouraged to leverage the power of external software libraries, in particular to ease the integration of their existing quantum solutions into \QCIConnect. By using separate repositories, cross-conflicts between dependencies of different extensions are avoided.
    \item \emph{Granular access control for internal and external users to enable \IP protection:} While the \SDK as a basis for all \AppLib extensions is made available to all interested users, app developers, in particular from industry partners, are enabled to restrict access to their applications to selected user groups.
    Partner libraries can enforce tiered access for different user groups:
    \begin{enumerate}
        \item \emph{Web Demo:} Access to running the application through the web interface of \QCIConnect. This will be possible by letting the application be verified by an \ALF server administrator, who can then add it to the portfolio hosted and made available through the \GUI.
        \item \emph{Local Installation:} Installation of compiled application code to run from scripts. The partner can provide the installation themselves through established package managers, on their repository page, or on their own websites. They can also contact the \QCIConnect service mail to inquire about providing the link on the \QCIConnect webpage for higher outreach.
        \item \emph{Code Access:} Full access to application's source code. This can be managed by the owner of the library repository themselves.
    \end{enumerate}
\end{enumerate}

\begin{figure}[tb]
    \centering
    \def\svgwidth{\linewidth}
    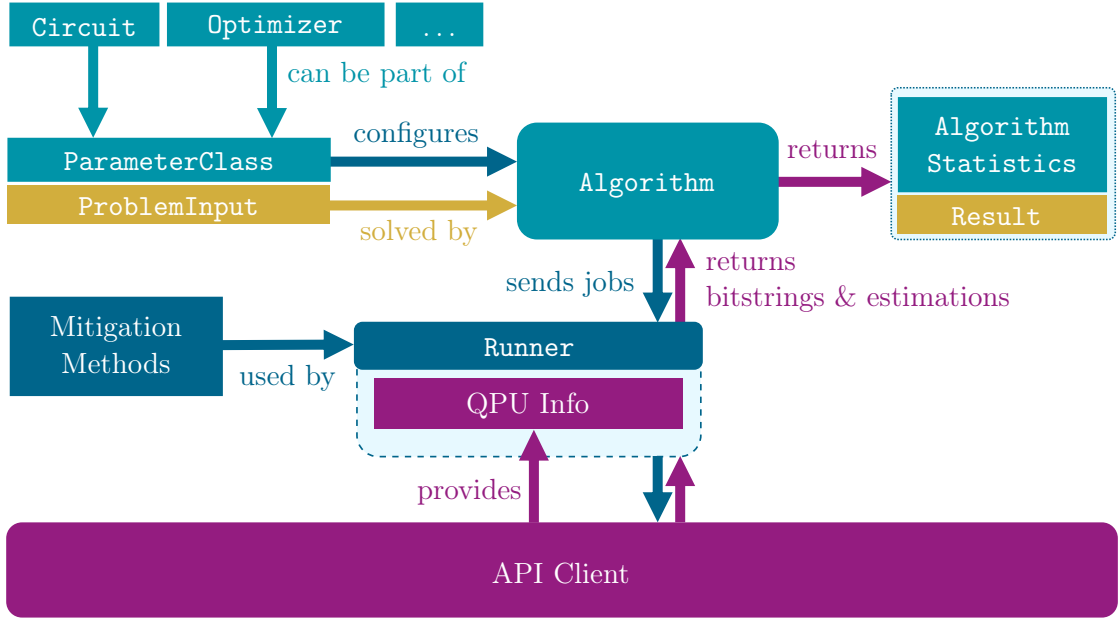
    \caption{Sketch of relations between ALF's main interface classes. 
    An algorithm is provided parameters, which tune its behavior, and a \texttt{Runner} that contains all information about the QPU and mitigation methods and is used for any quantum computation steps. 
    An algorithm that is supplied a parametrization and a \texttt{Runner} can take a \texttt{ProblemInput} subclass instance and return instances of both a \texttt{Result} subclass (containing information about the problem solution) and a \texttt{AlgorithmStatistics} subclass (containing information about the algorithm's performance).}
    \label{fig:alf-structure}
\end{figure}

Partners can use the \emph{Template \AppLib} as a basis to start implementing their own \AppLib. This template is a minimal repository that contains the necessary files which need to be filled with the application logic and in particular the configuration files for a web demonstration.
In addition, the documentations for \ALF and \AppLib contain tutorials that showcase how to implement applications in \ALF and how to use the reference applications in \AppLib.
\autoref{fig:alf-structure} provides an overview of the main interface classes of \ALF and how they relate to one another.

In practice, implementing an application based on \ALF contains the following steps:
\begin{enumerate}
    \item \emph{Define Problem Input:}
    \ALF uses inheritance to describe which problem classes an application can solve, such that for a given problem valid applications can be found and vice-versa for a given application examples of valid problems can be filtered for showcases.
    Therefore, the app developer can either choose an existing subclass of \ALF's \texttt{ProblemInput} that his application can solve, or implement a dataclass that either inherits directly from \ALF's \texttt{ProblemInput} class or from a more specialised subclass. 
    \item \emph{Define Result:} Similar to the problem definition, the application's result class can sit anywhere in the inheritance tree of \texttt{Result}. In addition, application may define custom subclasses of \texttt{AlgorithmStatistics} to track custom benchmark statistics for their application, such as fine-grained runtime information, iteration counts, and convergence metrics.
    \item \emph{Define Algorithm Parameters:}
    Choose an existing class or implement a dataclass that inherits from \texttt{ParameterClass} and contains all parameters which can be changed by the user to tune the application.
    \item \emph{Implement Core Logic:} Depending on complexity of the application:
    \begin{enumerate}
        \item \emph{Classical-quantum-classical algorithm as sketched in \autoref{fig:alf-app-workflow}}: Implement a subclass of \ALF's \texttt{SimpleAlgorithm}. This subclass must contain a preprocessing method that prepares two types of quantum jobs:
        \begin{enumerate}
            \item \emph{Sampling jobs} that contain quantum circuits whose bitstrings should be retrieved in the quantum computation step. These jobs also contain the number of shots to be used.
            \item \emph{Estimation jobs} that, in addition to circuits and number of shots, also contain a list of observables whose expectation values and, optionally, a hand-selected set of measurement bases for that estimation. In the quantum computation step, \ALF automatically computes the estimated expectation values from the bitstring measurements of these jobs.
        \end{enumerate}
        \item \emph{Hybrid algorithm}: Implement a subclass of \texttt{Algorithm}. This subclass can directly call the sampling and estimation routines of \ALF to seamlessly switch between quantum and classical steps in a distributed workflow.
    \end{enumerate}
    Notably, error mitigation need not be implemented directly as part of the algorithm, because the error mitigation techniques selected by the user are automatically applied through the sampling and estimation routines. 
    This ensures that developers can focus on the core logic of their algorithm.
\end{enumerate}

To also provide a web demonstration via the \GUI on the \QCIConnect webpage, 
the Template AppLib should be used as a reference point for defining:

\begin{enumerate}
    \item \emph{Metadata for App description:} the human-readable title, a short abstract to be displayed on the preview card, a Readme file for more detailed information and a thumbnail image.
    \item \emph{Input fields:} configuration files define which fields the web frontend should show (e.g., number of shots, number of steps, problem definition), what the default values are, which maximum or minimum values are allowed and which further description the web user should see. Users of the web demo typically do not require the same depth of customization as power users accessing the \SDK directly and it is recommended to keep the configuration options simple.
    \item \emph{Workflow:} the main entry point of the application package defines how the algorithm is initialised and executed and how the output files are created. The output files should include the raw results and a markdown file for rendering the results in the web frontend, which may include figures such as plots.
\end{enumerate}

Applications ready for a web demonstration should be hosted inside the DLR GitLab. The vetting process can be started by reaching out to the technical support of \QCIConnect with the URL and revision of the application.

\subsection{Existing AppLib Extensions}
\label{sec:sdk:applib-extensions}

One example of an extension library is the \emph{quantum optimization} \AppLib \package{alqo}. It builds on top of \quark and \quapps, 
two packages that have been developed at \DLR~\SC for many years already and that are steadily extended as part of the open source package collection ``DLR-SC Quantum Computing Software''~\cite{dlrsc2026repo}.
They automatise the pre-processing of optimization problems for quantum computation. Specifically, \quark handles the mapping of arbitrary polynomial optimization problems to Ising/QUBO problems~\cite{lobe2023quark, lobe2026quarksoftware}, 
which form the base line for several quantum optimization algorithms~\cite{lucas2014ising}. Additionally, \quapps provides a library of pre-implemented and adaptable optimization problems~\cite{lobe2026quappssoftware, windgaetter2025quantum}. 
Although the input to these problems is in general rather simple, e.g. just a set of weighted edges in case of the Travelling Salesperson problem, finding the optimal solution is computationally hard - at least in the classical world. 

Since all \quapps problems use the standardised interface of \quark, they can thus easily be inferred to different solvers.  
The \package{alqo} package integrates these two libraries into the \QCIConnect-SDK framework and provides multiple algorithm implementations for solving optimization problems with quantum computers, such as the Grover search algorithm or QAOA. Those, in particular, consist of multiple loops over algorithmic sub-procedures, which can easily be implemented individually based on the \ALF interfaces and then plugged together. Note that a lot of these algorithms are called ``hybrid'' but actually require well-organised distributed computational power rather than interventions within decoherence time.  

Due to the interchangeability of the \quapps problem definitions, each of the algorithms in \package{alqo} can be applied to each of the problems. 
This $n$-to-$m$ relation is a decisive difference to, for instance, quantum simulation where in general specific applications do also have very specific problem inputs. But due to the well-thought-out interfaces of \ALF, the implementation of different optimization algorithms can be done independently of the actual problem input. An application connecting the problem input with the algorithm can finally directly be run on the \QCIConnect hardware and obtained results be parsed back to give an answer to the original optimization problem. 

By complying with \ALF and providing accompanying material such as metadata, meaningful documentation and executable tutorials, we additionally have the opportunity to make the optimization applications available on the \QCIConnect web frontend. The graphical interface further lowers barriers for non-experts to experience quantum computing. 

With the instance data set that comes with \quapps~\cite{lobe2025quappsdataset} one has not only a baseline for showcasing the execution of an application in the full stack but actually also collected a full toolbox for executing exhaustive benchmarks. By keeping all elements in the stack but one, the exact influence of this component can be analysed in detail. This is a key element if practical quantum advantage shall be observed at some point. 

\medskip

The other main application area besides quantum optimization is \emph{quantum simulation}, being the original motivation for researching quantum computers.  
The \DLR institute \SC also comprises experts in quantum simulation who are currently working on actively implementing a respective application library. The diverse range of quantum simulation applications has in general a very close connection between problem inputs and respective algorithms, but at the same time also benefits from a compliance of hardware properties with the ones of the observed quantum system. 
Thus, to ensure an actual advantage, the layers in the software stack need to interlock as much as possible.

An example of such an application is \PITE~\cite{ejima2025probabilistic, ejima2025probabilisticsoftware}, which is used to find the ground state of quantum systems.
\PITE benefits greatly from flexible and easy-to-use state vector simulations, which enhance prototyping and tuning of the algorithm to new problems.
The implementation of \PITE motivated design choices for the future development of the \SDK to both ease the transition for users that come from different frameworks and to leverage the capabilities of simulators. This underscores the importance of application-aware software development to maximise usability.

\medskip

With the development of these application libraries, we have experienced that every research domain and the corresponding problems and applications have different requirements and thus pose different challenges on the framework. Therefore, \ALF cannot be established independently from but rather requires a co-development with the application developers: To provide a flexible but stable frame that can be operated and maintained for a long time and thus support as many quantum application developers as possible. 
    \section{Outlook}
\label{sec:outlook}

\QCIConnect is positioned as a hardware-agnostic quantum computing platform, whose modular architecture can scale and evolve with upcoming advances in quantum computing hardware. 
We envision several strategic directions for future developments to further increase the value for the quantum computing community.

The \API and \SDK are both open source, supporting a community-driven ecosystem of hardware, simulators, and applications.
They will continue to evolve with expanded functionality and improved documentation.
An AI assistant may help new users bring their use cases to \QCIConnect.
We're particularly focused on simplifying domain-specific application development while maintaining the platform's hardware-agnostic approach.

The platform's tool and application ecosystem will continue to grow.
New compilers and simulators, added through existing interfaces, will give \SDK users better tools to test and optimise their applications for specific hardware.
In particular, more applications based on the \ALF will be integrated, bringing advances from \DLR research institutes and from projects with application partners from industry to users of \QCIConnect.
An \ALF server will be hosted to support web demonstrations via the \GUI, enabling partners to easily showcase \IP-sensitive applications.
These applications can evolve into primitives to be used as part of larger workflows in the future.

While the \SDK already allows out-of-coherence time hybrid algorithms by alternating between quantum and classical computations on the client side, \QCIConnect can improve the workflow by introducing blocking sessions and classical computation on dedicated machines or \HPC systems. 
Beyond this, hybrid algorithms within coherence time, in particular classical feed forward for error correction, will become substantial tools as qubit numbers and quality improve.
These require fast classical computation, for instance through Field Programmable Gate Arrays.
To integrate such devices \QCIConnect will expand its orchestration capabilities to support more complex workflows to seamlessly integrate real-time feedback between quantum and classical components.
A major step towards that goal is the integration of a higher-level language beyond \OpenQASM and \QIR. 

Benchmark projects are planned to provide meaningful metrics for the full quantum software stack, from problem description to machine execution.
These metrics will enable systematic evaluation of workflows and co-design of software components to harness the full capabilities of the quantum hardware.

Another goal is a streamlined interface for using error mitigation through the full stack: at the current stage, the user running an application via \ALF is able to define a combination of mitigation methods that are performed on his machine. In the future, this should natively communicate with the physical layer to integrate with the compilation and mitigation steps provided by the vendors.

Interoperability with other software stacks, such as the \MQSS, can foster synergies and consolidate the strengths of both parties.
A consortium of different institutions under the lead of Fraunhofer FOKUS recently published a DIN SPEC for the ``Interface between quantum computer backends and software frameworks''~\cite{2025dinstandard}. With \DLR being part of the consortium, many of our choices learned from designing the \QCIConnect \API were incorporated into the SPEC and the \API will be continuously developed with the remaining specifications in mind.
Further homogenization with other software stacks can pave the way for long-term standards that benefit the full quantum computing ecosystem. 
With the further advance of the QC Next ``FullStaQD'' project~\cite{fraunhoferiao2026fullstaqdwebpage} and its accompanying satellite projects, 
further requirements on a full stack platform might be introduced, which we aim to adopt accordingly. 
In particular, the integration of the high-level programming language Eclipse Qrisp, can enable an even broader audience to develop their specific quantum applications.

Beyond these specific development directions, \QCIConnect benefits from continuous support: Regular updates that enhance features while maintaining stability, incorporation of user feedback and monitoring of emerging trends are the backbone of a healthy quantum computing platform.  
Through collaborations with partners from academia and industry to establish common standards, share best practices, and foster quantum solutions to industry pain points, \QCIConnect aims at becoming a cornerstone of Europe's emerging quantum computing ecosystem.
    
    \acknowledgement{The underlying research is part of the projects ‘Algorithms for quantum computer development in hardware-software codesign’ (ALQU), \url{https://qci.dlr.de/en/alqu},
and ‘Classical Integration of Quantum Computers’ (CLIQUE), \url{https://qci.dlr.de/en/clique}, 
which were made possible by the DLR Quantum Computing Initiative (QCI) and the German Federal Ministry of Research, Technology and Space (BMFTR). 

Through an industry call, d-fine and PlanQC have been chosen as industry partners for the design, conception and development of \QCIConnect, with \DLR focusing its resources on strategic project lead, alignment with hardware providers, integration into the \DLR ecosystem, and the development of the \QCIConnect \SDK.

The authors affiliated with d-fine thank their internal quantum computing Q-lab working group, \url{https://www.d-fine.com/en/solutions/q-lab/}, for stimulating discussions. 

The authors affiliated with DLR thank Michael Epping and Benedikt Fauseweh for the acquisition of the funding for this project, Michael Epping and Yoshinta Wied for their support during the tender calls, and Thorsten Sommer for the planning of the initial internal IT infrastructure. 
}
    
    \printbibliography

\end{document}